\documentstyle[emulateapj]{article}
\newcommand{\ts}{\thinspace}

\begin{document}

\title{NEW NEAR-INFRARED SPECTROSCOPY OF THE HIGH REDSHIFT QUASAR
       B{\ts}1422+231 AT $z${\ts}={\ts}3.62}

\author{\sc Takashi Murayama\altaffilmark{1}
        and Yoshiaki Taniguchi\altaffilmark{1}}
\affil{Astronomical Institute, Tohoku University,
                Aoba, Sendai 980-8578, Japan}
\affil{Electronic mail: murayama@astr.tohoku.ac.jp,
                            tani@astr.tohoku.ac.jp}

\author{\sc Aaron S.~Evans\altaffilmark{1}}
\affil{California Institute of Technology, 105-24,
       Pasadena, CA 91125}
\affil{Electronic mail: ase@astro.caltech.edu}

\author{\sc D.~B.~Sanders and K.-W.~Hodapp}
\affil{Institute for Astronomy, University of Hawaii,
      2680 Woodlawn Drive, Honolulu, HI 96822}
\affil{Electronic mail: sanders@ifa.hawaii.edu, hodapp@ifa.hawaii.edu}

\author{\sc Kimiaki Kawara and Nobuo Arimoto}
\affil{Institute of Astronomy, The University of Tokyo,
                 2-21-1 Osawa, Mitaka, Tokyo 181-8588, Japan}
\affil{Electronic mail: kkawara@mtk.ioa.s.u-tokyo.ac.jp,
                        arimoto@mtk.ioa.s.u-tokyo.ac.jp}

\altaffiltext{1}{Visiting Astronomer of the University of Hawaii
                 2.2 meter telescope.} 

\authoremail{murayama@astr.tohoku.ac.jp}

\begin{abstract}
We present new near-infrared (rest-frame UV-to-optical) spectra of the
high redshift, gravitationally lensed quasar B{\ts}1422+231 ($z=3.62$).
Diagnostic emission lines of \ion{Fe}{2}, \ion{O}{3}$\lambda$5007,
and H$\beta$,
commonly used to determine the excitation, ionization, and chemical
abundances of radio-quiet and radio-loud quasars, were detected.  Our new
data show that the ratio \ion{Fe}{2}(UV)/H$\beta${\ts}={\ts}18.1$\pm$4.6 and
\ion{Fe}{2}(optical)/H$\beta${\ts}={\ts}2.3$\pm$0.6 are
higher than those reported by Kawara et al.\ (1996)
by factors of 1.6 and 3.3, respectively, although the
ratio [\ion{O}{3}]5007/H$\beta${\ts}={\ts}0.19$\pm$0.02 is nearly the
same between the two measurements.
The discrepancy of the line flux ratios between the
measurements is likely due to improved data and fitting procedures rather
that to intrinsic variability.  While approximately half of the high-$z$
quasars observed to date have much more extreme \ion{Fe}{2}(optical)/H$\beta$
ratios, the line ratio measured for B{\ts}1422+231 are consistent with the
observed range of \ion{Fe}{2}(optical) ratios of low-$z$ quasars.
\end{abstract}


\section{INTRODUCTION}

Since near-infrared (NIR) spectroscopy of high-redshift ($z > 2$) quasars
provides information about their rest-frame optical spectroscopic properties,
it is now possible to systematically compare the spectroscopic  properties
(e.g., excitation, ionization, and chemical abundances) of 
high-$z$ and low-$z$ quasars.
Although  the first NIR spectroscopic  observations of high-$z$ quasars
were made nearly two decades ago (Hyland, Becklin, \& Neugebauer 1978;
Puetter et al.\ 1981; Soifer et al.\ 1981), high-quality
NIR spectra have been obtained for only $\sim${\ts}10 high-$z$ quasars to date
(e.g., Carswell et al.\ 1991;
\markcite{Hill93}Hill et al.\ 1993; \markcite{Elston94}Elston et al.\  1994;
Kawara et al.\ 1996 [hereafter K96]; Taniguchi et al.\ 1997a; 
Murayama et al.\ 1998; see also Taniguchi et al.\ 1997b).
However, despite the relatively small number of objects observed, 
several very interesting properties of high-$z$ quasars have emerged.
In particular, Hill et al.\ (1993) and Elston et al.\ (1994; hereafter ETH)
suggested that unusually strong optical \ion{Fe}{2} emitter may be
common in the high-$z$ universe ($2<z<3.4$); to date,
five out of eight high-$z$ quasars
observed have very strong optical \ion{Fe}{2} emission with
EW(\ion{Fe}{2})/EW(H$\beta$)$\gtrsim 1$ (see Murayama et al.\ 1998).
By comparison, a much
smaller fraction of far-infrared (FIR) selected AGN ($L_{\rm FIR} \gtrsim
10^{11} L_{\sun}$) have strong optical \ion{Fe}{2} emission
(cf.\ L\'{\i}pari et al.\ 1993).

Recently, we obtained NIR spectra of the  radio-loud, flat-spectrum, high-$z$ quasar
B{\ts}1422+231 (Patnaik et al.\ 1992; Lawrence et al.\ 1992)
using the Mayall 4{\ts}m telescope at Kitt Peak National Observatory (KPNO) (K96). 
Although this quasar is at $z=3.62$, its optical magnitude is
sufficiently bright ($m_{\rm r}$ =15.5; Yee \& Bechtold 1996), due to 
gravitational lensing (Patnaik et al.\ 1992; Lawrence et al.\ 1992),
to allow it to be studied by NIR spectroscopy.
The NIR spectra show that the flux ratio
$F(\mbox{\ion{Fe}{2} }{\ts}\lambda\lambda\mbox{4434--4684})/F({\rm H}\beta)$ 
is much less than that of the other high-$z$ quasars (K96), 
and in fact is  similar to those of radio-loud, flat-spectrum, low-$z$ quasars 
with ``normal'' optical \ion{Fe}{2} emission.
This suggested that high-$z$ quasars may exhibit a range of values of
$F(\mbox{\ion{Fe}{2} }{\ts}\lambda\lambda\mbox{4434--4684})/F({\rm H}\beta)$ 
similar to what has been observed for low-$z$ quasars.
In order to confirm this result, we have obtained new NIR spectra using
the University of Hawaii (UH) 2.2{\ts}m telescope. In this paper, 
we present our new NIR spectroscopy of B{\ts}1422+231 and compare it with our previous
measurements.

\section{OBSERVATIONS AND DATA REDUCTION}

We observed B{\ts}1422+231 on 1996, April 7 (UT) using the K-band
spectrograph (KSPEC; \markcite{Hodapp94}Hodapp et al.\ 1996)
at the Cassegrain focus (f/31) of the UH 2.2{\ts}m
telescope in combination with the UH tip-tilt system
(\markcite{Jim98}Jim et al.\ 1998).
The cross-dispersed echelle design of KSPEC
provided simultaneous coverage of the entire 
1--2.5 \micron{} wavelength region.
The projected pixel size of the HAWAII 1024 $\times$ 1024 array
was 0\farcs{}167 along the slit and $\simeq$ 5.6 \AA{} at
2 \micron{} along the dispersion direction.  We used a 0\farcs{}96 
wide slit oriented East-West and centered on the intensity peak
of component B (Patnaik et al.\ 1992) of B{\ts}1422+231 (see Figure 1).  
However, comparing the two images given in Figure 1, we find that
actual light from component B is 59{\ts}\% of the total measured flux
and the remaining 41{\ts}\% is associated with light contamination
from components A and C.  
Therefore, the flux of our spectra is 1.7 times as bright as real component B
flux.

Thirty exposures, each of 180 sec 
integration, taken under photometric conditions, were obtained by
shifting the position of the object along the slit
at intervals of 5\arcsec{} between each integration.
The total integration time was 5400 sec.
An A-type standard star, HD{\ts}106965
(\markcite{Elias82}Elias et al.\ 1982),
was observed for flux calibration.
Another A-type star, HD{\ts}136754
(\markcite{Elias82}Elias et al.\ 1982), was also observed
before and after observing
B{\ts}1422+231  to correct for atmospheric absorption.
Spectra of an incandescent lamp and an argon lamp were taken for
flat-fielding and wavelength calibration, respectively.
Typical widths of the spatial profiles of the
standard star spectra were $\sim${\ts}0\farcs{}5 (FWHM) throughout the night.
Note that our previous NIR spectroscopy of this quasar at KPNO 
was obtained using a 1.44 arcsec slit
under 1.4 -- 2.1 arcsec seeing conditions (K96).

\placefigure{fig-1}

Data reduction was performed with IRAF\footnote{%
Image Reduction and Analysis Facility (IRAF) is
distributed by the National Optical Astronomy Observatories,
which are operated by the Association of Universities for Research
in Astronomy, Inc., under cooperative agreement with the National
Science Foundation.} using standard procedures as outlined in 
\markcite{Hora96}Hora \& Hodapp (1996).
Sky and dark counts were removed by subtracting the average of
the preceding and following exposures, and then the resulting
frame was divided by a normalized dome flat.
The target quasar was not bright enough to trace
its position in each frame with sufficient accuracy. 
Therefore, we first fit the spectral positions of the standard star
spectra with a third-order polynomial function
which properly traces the echelle spectrum.
These fitting results were then applied to the quasar spectra.
Using this procedure, we extracted the quasar spectra with an aperture
of 3\arcsec{} which was determined to be the typical width
where the flux was $\sim${\ts}10{\ts}\% of
the peak flux along the spatial profile of the standard star.
In order to subtract the sky emission, we used the data just adjacent
to the 3\arcsec{} aperture.
The wavelength scale of each extracted spectrum was calibrated to an 
accuracy of 18 km s${}^{-1}$ at 1.1 \micron, 20 km s${}^{-1}$ at
1.6 \micron, and 19 km s${}^{-1}$ at 2.0 \micron, 
based on both the argon emission lines of the calibration lamp and
on the telluric OH emission lines.  
The spectral resolutions (FWHM) measured from the argon
lamp spectra were $\simeq$ 500 km s${}^{-1}$ at 1.1 \micron, $\simeq$
450 km s${}^{-1}$ at 1.6 \micron, and $\simeq$ 500 km s${}^{-1}$ at 2.0 \micron.
Finally, the spectra were median combined in each band.
Atmospheric absorption features were removed
using the spectra of the A-type star HD{\ts}136754
because A-type stars are best suited for correcting for atmospheric absorption
features.
However, since A-type stars inherently have hydrogen recombination
absorption lines (e.g., the Brackett series in $H$ and $K$ bands
and the Paschen series in $I$ and $J$ bands),
we removed these features using Voigt profile fitting before the correction. 
In order to check whether this procedure worked 
we also applied the same atmospheric correction for the spectra of M-type stars
whose data were obtained on the same night.
Comparing our corrected spectra of the M-type stars with their 
published spectra (Lan\c{c}on \& Rocca-Volmerange 1992),
we found that our correction procedure works appropriately.
Finally, in order to calibrate the flux scale, we used 
the spectrum of the standard star HD{\ts}106965 (A2, $K$=7.315)
divided by a 9000{\ts}K blackbody spectrum, which fits
the $J\!H\!K\!L$ magnitude of the standard
\markcite{Elias82} (Elias et al.\ 1982)
with only 1.2{\ts}\% deviation.
Photometric errors were determined to be $<${\ts}10{\ts}\%
over all observed wavelengths.

\section{RESULTS AND DISCUSSION}

Figure \ref{fig-2} shows the spectra of B{\ts}1422+231 (red line)
in the $I\!H\!K$ bands 
together with both our previous measurement at KPNO 
(blue line; K96) and 
the Large Bright Quasar Survey (LBQS) composite
spectrum shifted to $z=3.62$ (black dashed line; \markcite{Francis91}
Francis et al.\ 1991).  
Since the efficiency of KSPEC in the $J$-band is not high,
we have not used the $J$-band data in this paper.

\placefigure{fig-2}

Although our new measurement was made with a narrower slit (0\farcs{}96 wide)
than that used by K96 (1\farcs{}44 wide),
the $H$- and  $K$-band fluxes are slightly higher
than those of the previous measurement.
K96 tried to carefully correct for the effect
of seeing on the relative flux calibration among the spectral bands
because they could take only one band spectrum at a time.
However,
the time variation of the seeing in their observations made it difficult 
to perform the correction very accurately.
Since our new  $I$ to $K$-band NIR spectra were obtained simultaneously,
our new measurement should be more reliable than
that of K96.
Our new $I$-band spectrum detects \ion{C}{3}]$\lambda$1909 emission clearly.
Further, some unidentified emission features at 2000{\ts}\AA~
and 2080{\ts}\AA ~
as well as the dip at 2200{\ts}\AA~ in the rest frame, which are 
shown in the average spectrum of LBQS quasars (Francis et al.\ 1991),
are also seen.
The presence of [\ion{O}{3}]$\lambda$5007 emission has also been confirmed. 

The continuum flux of component B between 1330{\ts}\AA~ and
1380{\ts}\AA, which was obtained by Impey et al.\ (1996)
13 months before our observations, 
was assumed for the continuum level on the short
wavelength side of our spectra.
We applied the factor 1.7 to this continuum level 
in order to correct the contaminated light from component A and C for
our spectra (see previous section).   
The continuum level on the long wavelength side of our spectra 
was chosen by fitting a power-law
continuum plus a \ion{Fe}{2} template simultaneously to minimize the residual
between 4450{\ts}\AA~ and 4750{\ts}\AA~ and beyond 5100{\ts}\AA.  We used the Balmer continuum
template that was generated to approximate the emission-line strengths
seen in 0742+318 (see Figure 3d of Wills et al.\ 1985).  The \ion{Fe}{2}
emission profile of the very strong \ion{Fe}{2}-emitting low-ionization
BAL quasar PG{\ts}0043+039 (Turnshek et al.\ 1994)
was used as the \ion{Fe}{2} template.
Due to a relatively poor fit over the entire wavelength range
when using a single
template, two \ion{Fe}{2} templates were used: one for shortward of
3000{\ts}\AA \ and the other for longward of 3000{\ts}\AA.  
The power-law continuum that we derive is given by $f_\nu \propto \nu^{-0.88}$. 
This spectral index is steeper than the 
$f_\nu \propto \nu^{-0.54}$ power-law derived in K96. 
Yee \& Bechtold (1996) report that B{\ts}1422+231 had become brighter by 
0.12 mag during 13 months (see also Kundi\'c et al.\ 1997), 
but variability of 0.12 mag would change
the spectral index to $-0.88 \pm 0.08$ at most.
Therefore, it is unlikely that
the difference of the spectral index between K96 and our current work 
is due to the variability inherent in this quasar.
Thus, we conclude that the 
change in the calculated spectral index is due to
our improved absolute NIR spectrophotometry
by simultaneous observations of $I\!H\!K$ bands using KSPEC.

Figure 3 shows the rest frame spectrum of B{\ts}1422+231 with the power-law
subtracted and the best fit synthetic spectrum comprised of \ion{Fe}{2} and Balmer
continuum emission as well as other broad-lines. 
The fluxes, equivalent widths, and line widths
of the detected emission lines are summarized in Table 1
together with the previous measurements of K96.
Our new data show that the ratio \ion{Fe}{2}(UV)/H$\beta$ and
\ion{Fe}{2}(optical: $\lambda\lambda$3500--6000)/H$\beta$ are
higher than those of K96 
by  factors of 1.6 and 3.3, respectively.
These differences are mainly due to adopting the different power-law continuum
described above.
However,  our new \ion{Fe}{2}(optical)/H$\beta$ ratio
for B{\ts}1422+231 is still in the
observed range for low-$z$, radio-loud quasars
(Boroson \& Green 1992; see Taniguchi et al.\ 1997a; Murayama et al.\ 1998).
The [\ion{O}{3}]$\lambda$5007/H$\beta$ ratio is nearly the same 
between the two measurements.
We also note that there appears to be evidence for excess emission
at $\lambda <$ 2100{\ts}\AA ~ in the rest frame
(see the second panel of Figure 3).

Our new measurements yield a flux ratio of
\ion{Fe}{2}(UV)/\ion{Fe}{2}(optical) $\simeq$ 8.0.
Since Wills et al.\ (1985) give a range of
4 $<$  \ion{Fe}{2}(UV)/\ion{Fe}{2}(optical) $<$ 12
for low-$z$ quasars, B{\ts}1422+231 is typical of low-$z$ quasars
in this respect.
We also obtain a flux ratio of
\ion{Fe}{2}(optical: $\lambda\lambda$4484--4684)/H$\beta
\simeq 0.53$ which is significantly smaller than the range of values 
($\sim${\ts}1.5 -- 2) found by Hill et al.\ (1993) for quasars
with $z \simeq$ 2 -- 2.5{\ts}. 
This demonstrates that it is perhaps dangerous to attempt a measurement of 
the iron abundance using solely
optical \ion{Fe}{2} emission features as suggested by Wills et al.\ (1985).

Finally we speculate about the formation epoch of the host galaxy of
B{\ts}1422+231. Our new measurement has confirmed that
the \ion{Fe}{2}(total)/H$\beta$
ratio for B{\ts}1422+231 is higher-than-normal with respect to
what is found for low-$z$ quasars.
Since the low-$z$ quasars are believed to be associated with nuclei
of massive galaxies, their chemical abundances are expected to be higher than
or roughly equal to solar.  
Therefore, the observed higher \ion{Fe}{2}(total)/H$\beta$ ratio suggests that 
the iron abundance of B{\ts}1422+231 is at least comparable to the solar value.
If this is the case, it is expected that the majority of iron would come from
Type Ia supernovae.
Yoshii, Tsujimoto, \& Nomoto (1996) derived $\sim$\ 1.5 Gyr
for the lifetime of SN Ia progenitors from an analysis of
the O/Fe and Fe/H abundances in solar neighborhood stars.
If the Fe enrichment started at 1.5 Gyr after
the onset of the first epoch of star formation,
the host galaxy of B{\ts}1422+231 would
have formed at $z \sim 9$ or earlier for $q_{0}$ = 0.0 and
$H_{0}$ = 75 km s$^{-1}$ Mpc$^{-1}$ (K96). 
\placefigure{fig-3}

\begin{table}
\dummytable\label{tbl-1}
\end{table}

\placetable{tbl-1}

\acknowledgments

We are very grateful to the staff of the UH 2.2 m telescope.
In particular, we would like to thank
Andrew Pickles for his technical support and assistance with the observations.
This work was financially supported in part by Grants-in-Aid for 
Scientific Research (Nos.\ 07044054 and 09640311)
from the Japanese Ministry of Education, Science, Sports, and Culture
and by the Foundation for Promotion of Astronomy, Japan.
TM is thankful for support from a Research Fellowship from the Japan
Society for the Promotion of Science for Young Scientists.
This research has made use of the NASA/IPAC Extragalactic Database
(NED) and the NASA Astrophysics Data System Abstract Service. 

\appendix

\begin{center}
{\bf Comments on continuum fitting}
\end{center}

In our spectral fitting procedure, we have used the global continuum
which was determined using the continuum from the rest-frame UV to optical.
However, local continua have often been used to measure
the optical \ion{Fe}{2}/H$\beta$
ratio for low-$z$ quasars because of the lack of rest-frame UV spectra
(e.g., Boroson \& Green 1992).
Since the adopted continuum affects the measurements of emission-line fluxes
(see Murayama et al.\ 1998), we examine such differences in the flux measurement
for the case of B{\ts}1422+231. In Figure 4, we  show the spectral fitting results
for both global continuum and the local continuum cases.
It is shown that the local continuum fit tends to give 
lower fluxes for the concerned emission lines; the measured fluxes for
H$\gamma$+[\ion{O}{3}]$\lambda$4363, H$\beta$, [\ion{O}{3}]$\lambda$ 5007,
and optical \ion{Fe}{2} are given in Table 1.
Comparing these fluxes with those measured using a global fit 
to the continuum, 
we find that the optical \ion{Fe}{2} flux based on the local continuum is four times
lower than that determined using a global continuum fit,
although the [\ion{O}{3}]$\lambda$5007
flux is nearly the same using both the local and global continuum fits.
Therefore, we suggest that previous measurements of
optical \ion{Fe}{2}/H$\beta$ ratios for low-$z$ quasars and
Seyfert nuclei may be underestimated by a factor of a few.
We also note that a local continuum fit should be used if one would
like to first compare new results with previous published values.

\clearpage


\figcaption[Murayama.fig1.ps]{%
The NIR ($K$-band) image of B{\ts}1422+231 taken
with the field viewer of KSPEC.
The weakest component D (0\farcs{}94 E and 0\farcs{}81 S
from component B) can be barely seen. 
In the lower panel, we show our 1 arcsec slit position.
\label{fig-1}
}

\figcaption[Murayama.fig2.ps]{%
Comparison of the observed-frame $I\!H\!K$ spectra of B{\ts}1422+231 (red line)
and both our previous measurement at KPNO 
(K96; blue line) and the LBQS composite spectrum of
Francis et al.\ (1991;  black dashed line).
\label{fig-2}
}

\figcaption[Murayama.fig3.ps]{%
The upper panel gives the rest frame spectrum of 
B{\ts}1422+231 and the same spectrum with the power-law continuum subtracted,
together with the  power-law continuum and a synthetic spectrum comprised 
of \ion{Fe}{2} and Balmer continuum emission plus other emission lines.
Note that the rest frame spectrum was derived by dividing the observed
frame spectrum by ($1+z$) after converting observed wavelengths into the rest-frame 
wavelengths. The spectrum shown in the second panel is the extracted
\ion{Fe}{2} emission of B{\ts}1422+231 together with
the best-fit \ion{Fe}{2} template.
The third and bottom panels show the best-fit templates of the Balmer
continuum and other emission lines respectively.
\label{fig-3}}

\figcaption[Murayama.fig4.ps]{%
The profile fit of the $K$-band spectrum.
The upper panel shows the result using the global power-law continuum
while the lower panel shows the result using the local continuum.
Note that the flux is multiplied by ($1+z$) for deredshifting.
\label{fig-4}}

\end{document}